\documentclass[journal,article,submit,moreauthors,pdftex]{Definitions/mdpi}

\renewcommand{\linenumbers}{}

\firstpage{1} 
\pubvolume{1}
\issuenum{1}
\articlenumber{0}
\pubyear{2023}
\copyrightyear{2023}
\datereceived{ } 
\dateaccepted{ } 
\datepublished{ } 

\hreflink{https://arxiv.org/abs/2309.01864}

\newcommand{\Ga}{\Gamma}
\newcommand{\de}{\delta}
\newcommand{\ep}{\varepsilon}
\newcommand{\la}{\lambda}

\newcommand{\pa}{\partial}

\newcommand{\be}{\begin{equation}}
\newcommand{\ee}{\end{equation}}
\newcommand{\bea}{\begin{aligned}}
\newcommand{\eea}{\end{aligned}}

\newcommand{\bml}{\begin{subequations}}
\newcommand{\eml}{\end{subequations}}

\newcommand{\bbm}{\begin{bmatrix}}
\newcommand{\ebm}{\end{bmatrix}}
\newcommand{\bvm}{\begin{vmatrix}}
\newcommand{\evm}{\end{vmatrix}}

\Title{Second-Order Transport Coefficients in Neutron Star Mergers}

\TitleCitation{Title}

\Author{Yumu Yang $^{1,}$*\orcidA{}, Mauricio Hippert\orcidB{} $^{1}$, Enrico Speranza\orcidC{}  $^{2}$ and Jorge Noronha\orcidD{} $^{1}$}

\AuthorNames{Yumu Yang, Mauricio Hippert, Enrico Speranza and Jorge Noronha}

\AuthorCitation{Yang, Y.; Hippert, M.; Speranza, E.; Noronha, J.}

\address{%
$^{1}$ \quad Illinois Center for Advanced Studies of the Universe \& Department of Physics, University of Illinois Urbana-Champaign, Urbana, IL 61801, USA \\
$^{2}$ \quad Theoretical Physics Department, CERN, 1211 Geneva 23, Switzerland}

\corres{Correspondence: yumuy2@illinois.edu}

\abstract{In neutron stars, flavor-changing weak interactions determine the equilibrium fraction of protons over neutrons. In binary neutron-star mergers, violent changes in density modify this equilibrium value at timescales of milliseconds, comparable to those required for weak interactions to take place. As a result, the fraction of protons evolves out of phase with the density oscillations, giving rise to irreversible processes. The corresponding shift in pressure leads to dissipative work that can be modeled as an effective bulk-viscous correction. In this work, we derive the relevant equations of motion of Israel-Stewart hydrodynamics within this context. Using a toy model, we compute the second-order transport coefficients. Finally, we comment on the use of a realistic equation of state. Our results are expected to be useful for the study of viscous effects in numerical simulations of binary mergers.}


\begin{document}

\section{Introduction}

Neutron star mergers present a unique opportunity to explore the out-of-equilibrium properties of dense matter \cite{Alford:2017rxf}. Oscillations in temperature and density during these mergers can push the system away from beta equilibrium \cite{Hammond:2021vtv, Most:2021ktk}. 
As a response to the departure from detailed balance, weak interaction rates dynamically adjust the proton fractions to restore beta equilibrium. Because the oscillations occur at the same timescale as the weak interactions, the proton fraction evolves out of phase with changes in baryon density. This phase difference leads to dissipative work \cite{Alford:2018lhf, Alford:2021ogv, Alford:2023gxq}. 

In linear response theory, the dissipative effects associated with changing particle fractions can be modeled as an effective bulk-viscous correction to pressure \cite{Sawyer:1989dp, Haensel:1992zz}. Recent work in Ref.\ \cite{Gavassino:2020kwo} (see also \cite{Celora:2022nbp, Camelio:2022ljs, Camelio:2022fds}) showed that near beta equilibrium, the correction to the beta equilibrium pressure follows the Israel-Stewart equation \cite{Israel:1979wp}. The associated transport coefficients are determined by the equation of state (EoS) for dense matter in beta equilibrium, along with weak interaction rates \cite{Gavassino:2020kwo}.

In this work, we will show the derivation and computation of transport coefficients of the Israel-Stewart theory for neutron star mergers. 

\section{Israel-Stewart theory from chemical imbalance}
\label{IS derivation}

We consider neutron star matter at a low enough temperature that the neutrino mean free path is larger than the radius of the star \cite{Alford:2018lhf}. This system is neutrino-transparent, and, at sufficiently low densities, it is composed mostly of protons $p$, neutrons $n$, and electrons $e$. In this case, flavor equilibration occurs via the direct and modified Urca processes,
\begin{gather}
    n \to p + e^- + \Bar{\nu}_e ,\\
    p + e^- \to n + \nu_e ,\\
    n + X \to p + e^- + \Bar{\nu}_e + X ,\\
    p + e^- + X \to n + \nu_e + X,
\end{gather}
where $X$ is a spectator nucleon, a neutron or a proton. 

For a neutrino-transparent system, the beta equilibrium condition can be written via the detailed balance principle:
\be
\label{eq:beta_eq}
\mu_n = \mu_p + \mu_e ,
\ee
where $\mu_n$, $\mu_p$, and $\mu_e$ are the chemical potentials for neutrons, protons, and electrons, respectively. The deviation from equilibrium can be characterized by $\delta \mu = \mu_n - \mu_p - \mu_e$, so $\de \mu = 0$ denotes beta equilibrium. 

We assume the energy-momentum tensor to (formally) be that of an ideal fluid,
$
T^{\mu\nu} = (\ep + P)u^\mu u^\nu - P g^{\mu\nu},
$ 
where $\ep$ is the energy density, and $P$ is the total pressure. Neglecting energy loss from neutrino emission, we can write the conservation of energy, momentum, and baryon number, as well as the evolution of the electron fraction, respectively, as 
\begin{equation}
\nabla_\mu\left((\varepsilon+P) u^\mu u^\nu - P g^{\mu\nu}\right)= 0, \qquad \nabla_\mu (n_B u^\mu) = 0, \qquad
u^\mu \nabla_\mu Y_e = \frac{\Gamma_e}{n_B}, 
\end{equation}
where $n_B$ is the baryon density, $Y_e$ is the electron fraction, and $\Ga_e$ is the reaction rate, which we approximate to be linear in $\de\mu$ such that $\Gamma_e = \la \delta \mu$. 

In the linear regime, we can expand the pressure with respect to $\de\mu$ around $\beta$ equilibrium,
$ 
P(\varepsilon, n_B, \delta \mu) = P|_{\delta \mu = 0} + P_1 \delta \mu ,
$ 
where $P_1 = \left.\frac{\partial P}{\partial \delta \mu}\right|_{\varepsilon, n_B, \delta \mu = 0}$ and define the bulk scalar to be $\Pi = P_1 \delta \mu$. Furthermore, we can define partial equilibrium states with all non-conserved variables. Writing $\delta \mu = \delta \mu(\varepsilon, n_B, Y_e)$ for one such state, we have
\begin{equation}
\label{beta chain}
    u^\mu \nabla_\mu \delta \mu = \frac{\partial \delta \mu}{\partial \varepsilon}\bigg|_{n_B, Y_e} u^\mu \nabla_\mu \varepsilon + \frac{\partial \delta \mu}{\partial n_B}\bigg|_{\varepsilon, Y_e} u^\mu \nabla_\mu n_B + \frac{\partial \delta \mu}{\partial Y_e}\bigg|_{\epsilon, n_B} u^\mu \nabla_\mu Y_e.
\end{equation}
With the relations above, we can derive an Israel-Stewart-like equation \cite{Gavassino:2020kwo}
\begin{equation}
\label{eq:ISfinalform}
    \tau_{\Pi} u^\mu \nabla_\mu \Pi + \delta_{\Pi\Pi}\theta\, \Pi + \Pi = - \zeta \theta ,
\end{equation}
where the transport coefficients are given by
\begin{subequations}
\label{transport_coeff}
\begin{align}
    \tau_\Pi &= - \frac{n_B}{\la} \left.\frac{\pa Y_e}{\pa \de\mu}\right|_{\ep, n_B}, \\
    \zeta &= \tau_\Pi P_1 n_B \left( \left.\frac{\partial \delta \mu}{\partial n_B}\right|_{\varepsilon, Y_e} + \left.\frac{\varepsilon + P}{n_B}  \frac{\partial \delta \mu}{\partial \varepsilon}\right|_{n_B, Y_e} \right) ,\\
    \delta_{\Pi\Pi}&=\frac{\tau_{\Pi}} {P_1}\left[ \frac{\partial P_1}{\partial \varepsilon}\bigg|_{n_B}(\varepsilon+P|_{\delta \mu=0}) + \left.\frac{\partial P_1}{\partial n_B}\right|_\varepsilon n_B \right] .
    \end{align}
\end{subequations}

\subsection{AC bulk viscosity from Israel-Stewart theory}

Some studies explore periodic perturbations, introducing frequency dependence to bulk viscosity \cite{Harris:2020rus, Sawyer:1989dp, Sad:2009hba}. A frequency-dependent viscosity can be extracted from the Israel-Stewart equations via a Kubo formula. We can define the Green's function arising from the linear correction to $T^{\mu\nu}$ in response to a metric perturbation $\delta g_{\mu\nu}\propto e^{-i\omega t}$. Transport coefficients can then be extracted from the Fourier-transformed retarded Green's function, via Kubo formulas.

To the metric perturbation corresponds an expansion rate
\begin{equation}
   \theta = \frac{ \partial_0 \sqrt{-\det(\eta_{\mu\nu} + \delta g_{\mu\nu})}}{\sqrt{-\det(\eta_{\mu\nu})}} = \frac{\partial_0\sqrt{1 - 2\,\eta^{\alpha\beta}\delta g_{\alpha\beta}}}{1} = -i\omega\,\eta^{\alpha\beta}\delta g_{\alpha\beta}.
\end{equation}
After some algebra, we can extract the AC bulk viscosity,
\begin{equation}
\label{eq:kubozeta}
    \zeta_{\textrm{AC}}(\omega) = \frac{1}{\omega} \operatorname{Im}\,G_R^\theta(\omega) = \frac{1}{1+\omega^2 \tau_\Pi^2}\,\zeta\,,
\end{equation}
which correctly recovers the DC bulk viscosity in the limit $\omega\to 0$. 

\section{Transport coefficients from a toy model of equation of state}
We now calculate the transport coefficients using the following toy model for the EoS
\be
P(\mu_B, \mu_I) = (\mu_B^2 + M \mu_I)^2 + \mu_B^2\mu_I^2, 
\ee
with the isospin-violating interaction,
\be
\frac{d n_I}{dt} = -g\mu_B^3\mu_I, 
\ee
where $M$ and $g$ are just some parameters. Using Eqs. \eqref{transport_coeff}, we can calculate the transport coefficients to be
\begin{equation}
    \tau_\Pi = \frac{2 (M^2 + 3\mu_B^2)}{3g\mu_B^2}, \qquad
    \zeta = \frac{4M^2\mu_B}{9g}, \qquad
    \delta_{\Pi\Pi} = \frac{4(M^2 + 3\mu_B^2)}{9g\mu_B^3}, 
\end{equation}
where we can see that the transport coefficients decrease with increasing reaction rates ($\propto g^{-1}$), and increase if we make the pressure more sensitive to $\mu_I$ by increasing $M$.

\section{Transport coefficients from realistic equations of state}

\begin{figure}[ht!]
    \centering
    \includegraphics[width=8cm]{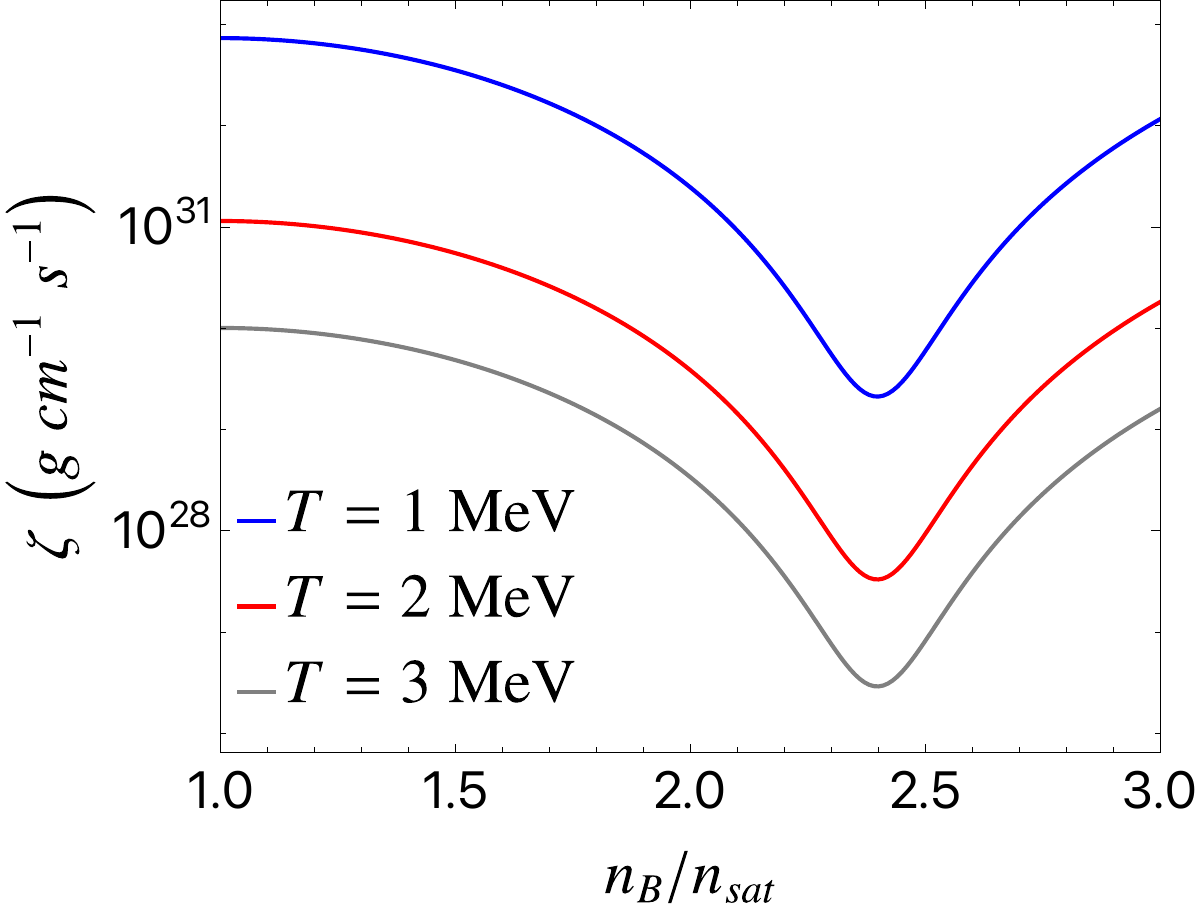} 
    \caption{Baryon density dependence of the bulk viscosity $\zeta$ at three different temperatures $T$. }
    \label{bulk}
\end{figure}

The transport coefficients can also be computed for realistic EoSs \cite{Yang:2023ogo}. We compute the transport coefficients for a set of relativistic mean field models with nucleons, interacting via  $\rho$, $\omega$, and $\sigma$ mesons, and electrons. These models are suitable for describing dense matter ranging from one nuclear saturation density to three nuclear saturation densities. In Figure.~\ref{bulk}, using as an example one such EoS, we show the bulk viscosity coefficient as a function of the baryon density normalized by the nuclear saturation density $n_{sat}$, at three different temperatures $T$. The bulk viscosity increases as temperature increases, and it has a nonlinear dependence on the baryon density. The bulk viscosity reaches a local minimum at around 2.4 saturation densities at all three temperatures.

\section{Conclusions}
We showed the derivation of Israel-Stewart hydrodynamics for systems in partial beta equilibrium. The procedure is general enough to be applied to all kinds of EoSs for neutron star matter. In this work, we considered $npe$ matter in the neutrino-transparent regime. In order to calculate the bulk transport coefficients, we used a toy model as an illustration and commented on realistic EoSs. We also showed how AC bulk viscosity can be obtained from metric perturbation. As an extension of this work, we also developed a more complete, fully nonlinear \cite{Gavassino:2023xkt} treatment for realistic EoSs in \cite{Yang:2023ogo}.

\vspace{6pt}

\funding{JN is partly supported by the U.S. Department of Energy, Office of Science, Office for Nuclear Physics under Award No. DE-SC0023861. MH and JN were partly supported by the National Science Foundation (NSF) within the framework of the MUSES collaboration, under grant number OAC-2103680. This research was partly supported by the National Science Foundation under Grant No. NSF
PHY-1748958. ES has received funding from the European Union’s Horizon Europe research and innovation program under grant agreement No. 101109747.}

\dataavailability{Data is contained within the article or supplementary material.}

\acknowledgments{We thank J.~Noronha-Hostler, L.~Gavassino, and M.~Alford for enlightening discussions, L.~Brodie and A.~Haber for providing assistance with the QMC-RMF1 EoS, and C.~Conde for providing us with the QLIMR module developed within the MUSES collaboration, which we have used to solve the TOV equation.}

\conflictsofinterest{The authors declare no conflict of interest.} 

\begin{adjustwidth}{-\extralength}{0cm}

\reftitle{References}

\bibliography{references,notInspire}

\PublishersNote{}
\end{adjustwidth}
\end{document}